\documentclass[12pt]{article}
\usepackage{graphicx,amssymb,amsmath,epsfig}
\usepackage{color}
\usepackage{subfigure}
\begin{document}

\begin{center}
{\large \bf Area products for ${\cal H}^{\pm}$ \footnote{ ${\cal H}^{+}$  
and ${\cal H}^{-}$ denote event horizon and Cauchy horizons} in AdS space}
\end{center}

\vskip 5mm

\begin{center}
{\Large{Parthapratim Pradhan\footnote{E-mail: pppradhan77@gmail.com}}}
\end{center}

\vskip  0.5 cm

{\centerline{\it Department of Physics}}
{\centerline{\it Vivekananda Satavarshiki Mahavidyalaya}}
{\centerline{\it West Midnapur, West Bengal~721513, India}}

\vskip 1cm

\begin{abstract}
We derive the thermodynamic products particularly area (or entropy) products of ${\cal H}^{\pm}$ for certain
class of black holes in AdS space. We show by explicit and exact calculations that more complicated function of event 
horizon area and Cauchy horizon area is indeed mass-independent. This mass-independent quantities indicate 
that they \emph{could turn out to be an ``universal'' quantity} provided that they depends only on the quantized 
angular momentum, quantized charges, and cosmological constant respectively. Furthermore, these area (or entropy) product 
relations for several class of black holes in AdS space gives us strong indication to understanding the nature 
of non-extremal black hole entropy (both inner and outer) at the microscopic level.  Moreover, we compute the 
 Penrose 's famous \emph{Cosmic Censorship Inequality} (which requires Cosmic-Censorship hypothesis) for these class of 
black holes in AdS space . Local thermodynamic stability has been discussed for these black holes and under certain 
condition certain black holes displayed second order phase transition.

\end{abstract}

\section{Introduction}
It has been explicitly examined that the thermodynamic products for Reissner Nordstr{\o}m (RN) BH, Kerr BH and 
Kerr-Newman (KN) BH \cite{ah09}: a simple area product of ${\cal H}^{\pm}$ is sufficient to draw a conclusion 
that the product of area (or entropy) is an \emph{universal} quantity. 
For instance, the product of inner or Cauchy horizon area (${\cal A}_{-}$) and outer or event horizon area (${\cal A}_{+}$)  
of standard four dimensional Kerr-Newman (KN) BH should read \cite{ah09}
\begin{eqnarray}
{\cal A}_{-}{\cal A}_{+} &=& 64 \pi^2 J^2+16\pi^2Q^4 ~.\label{pKN}
\end{eqnarray}
This expression indicates that the inner and outer horizon area product does depend on the quantized charges and 
quantized angular momentum but indeeds independent of the ADM (Arnowitt-Deser-Misner) mass of the BH space-time.
Since this only single area product relation defined on the horizon and independent of the ADM mass parameter thus 
it should be treated as an ``universal'' relation or ``universal'' quantity. Alternatively, we could say a unique 
product relation/formula only involving horizon quantities could turn out to be universal, but this is only a 
necessary condition. It should be emphasized throughout the manuscript that the ``universal'' term used for 
particular relations that are \emph{mass-independent}.  

The first motivation comes from the work of Visser \cite{mv13}. In this work, the author first demonstrated that  
in case of Schwarzschild-de Sitter (Kottler) BH in four dimensions, the area product of cosmological horizon 
and event horizon is not mass independent. Analogously, for RN-AdS \cite{mv13} BH in four dimensions it has been 
proved that the product of Cauchy horizon area and event horizon area is not explicitly mass independent. This means that 
the mass independence of two physical horizons area ( which are calculated perturbatively) is not generic \cite{mv13}. 

Since one can not derive a simple area product relation as in Eq. (\ref{pKN}), instead one could derive some complicated 
function of event horizon area and Cauchy horizon area  for above described spacetime that might be mass-independent and 
this mass-independent relation could turn out to be an ``universal'' quantity. Again these mass independent formulae of 
two physical horizon areas in spherically symmetry cases are intimately related to the  quasi-local quantities which could 
potentially hold in a wider setting. For example, when we have set $Q=0$ in Eq. (\ref{pKN}) one obtains the area product 
relation for Kerr BH in this case a simple area product is sufficient and it only depends on the quantized angular momentum 
of the BH, as can easily be seen from the exact solution.

It has been extended by Hennig \cite{jh} that for KN-AdS BH, some complicated function of inner and outer horizon area is 
indeed mass independent and that could turn out to be universal. Very recently, we have derived a functional relation of 
${\cal H}^{+}$ area and ${\cal H}^{-}$ area for a regular Ay\'{o}n-Beato and Garc\'{i}a BH \cite{ppgrg}. It suggests  
that some complicated function of horizons area that turns out to be universal. But it is not a simple area product of 
horizon radii as in RN BH, Kerr BH and KN BH. These has been very fascinating topic in recent years in the 
GR (General Relativity) community \cite{ah09} as well as in the String theory community \cite{cgp11} (see also \cite{pp14,plb}).

The second  motivation comes from the work of Cveti\v{c} et al. \cite{cgp11}, where the authors argued that 
if the \emph{cosmological constant is quantized} then the area (or entropy) product relations for general rotating multicharge 
BH in four and higher may provide a ``looking glass for probing the microscopics of general BHs''. Thus it is quite natural 
to study the area (or entropy) products after introduction of the cosmological parameter.  

However in this work, we intend to extend our study for several interesing  class of BHs in \emph{AdS space} for different class 
of theory. We derive the formulae for these BHs that involve the area (or entropy) relation in terms of BH horizons and BH 
parameters with a focus on ``universal" relations that are mass-independent. These mass independent formulae gives strong 
indication towards  they could turn out to be an ``universal'' quantity.

Phase transition \cite{hp83,david12} is an important phenomena in BH thermodynamics where the BH changes its phase from stability 
region to instability region. In case of Schwarzschild-AdS BH, Hawking and Page \cite{hp83} in 1983 first demonstrated that 
there exists first order phase transition between small and large BH. On the other hand, Kubiz\v{n}\'{a}k and Mann \cite{david12} 
first showed  there occurs second order phase transition for charged AdS BH in the extended phase space. Furthermore, we compute 
the specific heat for these BHs to examine whether they are locally stable or not. Finally, we examine whether these BHs possesses 
any kind of second order phase transition or not.

The third and final motivation comes from work of Penrose \cite{rp}. In this work, Penrose \cite{rp} had given an 
idea regarding the ``Cosmic Censorship Conjecture'' (CCC) which is an important issue 
in  the general theory of relativity that the total ADM (Arnowitt-Deser-Misner) mass $M$ of the Schwarzschild BH 
spacetime is related to the area ${\cal A}$  of the BH event horizon by the relation 
\begin{eqnarray}
M  &\geq&  \sqrt{\frac{{\cal A}}{16\pi}} ~.\label{scpi}
\end{eqnarray}
which is called the \emph{Cosmic Censorship Inequality} (CCI) or \emph{Cosmic Censorship Bound} (CCB) \cite{gibb05} and 
which is a necessary condition for Cosmic-Censorship hypothesis \cite{rp} (See \cite{bray,bray1,jang,rg,gibb99}).
We derive this inequality for these class of BHs. The physical significance of this inequality is that it states 
the lower bound on the mass for any time symmetric initial data set which fulfilled the Einstein equations with 
the negative cosmological constant. Additionaly, it is also fulfilled the dominant energy condition which possesses 
no naked singularities. Finally, we consider the thermodynamic properties in the extended 
phase space \cite{kastor09,david12} and in this framework, the cosmological constant is treated as thermodynamic pressure 
and its conjugate variable as a thermodynamic volume.

First, we consider the static and spherically symmetric solution in Ho\v{r}ava Lifshitz gravity with a cosmological 
constant. It is a UV complete theory of gravity. It is also a non-relativistic and re-normalizable theory of gravity 
\cite{ph9a,ph9b,ph9c}. It can be reduced to Einstein's general theory of relativity in the appropriate limit. This 
theory manifests a broken Lorentz symmetry in the UV cut-off region. 

The fact is that the  Horava-Lifshitz gravity is known to suffer from some perturbative instabilities in the 
IR limit \cite{blas,ch,li}, ultimately due to an extra mode which comes from the explicit breaking of general 
covariance. So one of the ways to fix this is to have Horava gravity emerging dynamically in the UV while preserving 
Lorentz invariance (or rather, not having a preferred foliation) in the IR. This is the proposal behind 
``covariant renormalizable gravity" proposed back in 2009 by Nojiri and Odintsov \cite{ss} and  the normal GR 
behaviour is naturally recovered in the IR limit, and the theory is stable \cite{gc}.

Secondly, we consider the BH solutions in massive gravity \cite{vegh}. In the holographic framework, it describes ``a clsss of 
strongly interacting quantum field theory with broken translation symmetry''. Massive gravitons are playing key role
in this theory. It has a ``Lorentz-breaking mass'' which is the alternative for ``spatial inhomogenities''. Due to this 
property, it breaks ``momentum conservation in the boundary field theory''. Whereas in Einstein's general theory of relativity, 
graviton is mass-less. Thus it is natural to have a question does any theory exists where the graviton mass is massive. Vegh 
\cite{vegh} obtains a BH solution in AdS space where graviton mass is massive. Thus it is interesing to study its thermodynamic 
properties and particularly focus on area (or entropy) product relations which are mass-independent.

Thirdly, we consider the another class of BH solution in AdS space where the fields are phantom. This phantom fields have 
negative energy density that will be discussed in Sec. 4. Finally, we consider another interesting class of BHs which are 
super-entropic. Their properties are discussed in Sec. 5.

\section{Thermodynamic Product Relations for Ho\v{r}ava Lifshitz BH in AdS Space:}

The static and spherically symmetric metric of Ho\v{r}ava Lifshitz BH in AdS Space \cite{mun} is 
given by 
\begin{eqnarray}
ds^2=-{\cal X}(r)dt^{2}+\frac{dr^{2}}{{\cal X}(r)}+r^{2}(d\theta^{2}+\sin^{2}\theta d\phi^{2}) 
~.\label{hlc1}
\end{eqnarray}
where the function ${\cal X}(r)$ is defined by
\begin{eqnarray}
{\cal X}(r) &=& 1+\left(\omega-\frac{2}{3}\Lambda \right) r^2-\sqrt{r\left[\omega\left(\omega-\frac{4}{3}\Lambda\right)r^3
+\beta\right]} ~.\label{hlc2}
\end{eqnarray}
where $\beta$ and $\Lambda$ are the integration constant and cosmological constant respectively.
Let us choose $\beta=4M\omega$, where $\omega$ is a free parameter that controls the UV 
characteristics of the theory. When $\Lambda=0$, we get back the solution of Kehagias-Sfetsos(KS) BH \cite{ks09} 
in Ho\v{r}ava Lifshitz (HL)\footnote{Within HL gravity, there are some BH solution which is non-asymptotically AdS spacetime
\cite{cai9}. But in our case the BH solution is asymptotically AdS spacetime.} 
gravity \cite{ph9a,ph9b,ph9c}. Again we recover the Schwarzschild BH in the IR limit 
$\omega \rightarrow \infty$. By inserting $\beta=4M\omega$, we can rewrite the function ${\cal X}(r)$ as
\begin{eqnarray}
{\cal X}(r) &=& 1+\left(1-\frac{2\Lambda}{3\omega} \right)\omega r^2-\omega r^2
\sqrt{1-\frac{4\Lambda}{3\omega}+\frac{4M}{\omega r^3}} ~.\label{hlc3}
\end{eqnarray}
The BH horizons could be find by setting  ${\cal X}(r)=0$ i.e. 
\begin{eqnarray}
4\Lambda^2 r^4+18\omega\left(1-\frac{2\Lambda}{3\omega}\right)r^2-36M\omega r+9 &=& 0 
~.\label{hlc4}
\end{eqnarray}
The roots of the equation can be find by applying Vieta's theorem:
\begin{eqnarray}
\sum_{i=1}^{4} r_{i} &=& 0 ~.\label{eq1}\\
\sum_{1\leq i<j\leq 4} r_{i}r_{j} &=& \frac{9\omega}{2\Lambda^2}\left(1-\frac{2\Lambda}{3\omega}\right) ~.\label{eq2}\\
\sum_{1\leq i<j<k\leq 4} r_{i}r_{j} r_{k} &=& \frac{9M\omega}{\Lambda^2} ~.\label{eq3}\\
\sum_{1\leq i<j<k<l\leq 4} r_{i}r_{j} r_{k}r_{l} &=&  \frac{9}{4\Lambda^2} ~.\label{eq4}
\end{eqnarray}

Case I:
Eliminating mass parameter, we obtain only single mass-independent relation in terms of 
two horizons:
\begin{eqnarray}
\frac{\frac{9}{4\Lambda^2}}{r_{1}r_{2}}-\left(r_{1}^2+r_{2}^2 +r_{1}r_{2}\right) &=& 
\frac{9\omega}{2\Lambda^2}\left(1-\frac{2\Lambda}{3\omega}\right) ~.\label{eq8}
\end{eqnarray}

One can rewrite this formula in terms of area ${\cal A}_{i} = 4\pi r_{i}^2, \, (i=1,2)$ and 
it should be
\begin{eqnarray}
\frac{\left(\frac{9\pi}{\Lambda^2}\right)}{\sqrt{{\cal A}_{1}{\cal A}_{2}}}-\frac{\left({\cal A}_{1}+{\cal A}_{2}+
\sqrt{{\cal A}_{1}{\cal A}_{2}}\right)}{4\pi} &=& 
\frac{9\omega}{2\Lambda^2}\left(1-\frac{2\Lambda}{3\omega}\right) ~.\label{eq9}
\end{eqnarray}
The other mass-independent formula becomes
\begin{eqnarray}
\sqrt{{\cal A}_{1}{\cal A}_{2}{\cal A}_{3}{\cal A}_{4}} &=& \frac{36\pi^2}{\Lambda^2} ~.\label{eq10}
\end{eqnarray}

Case II:
For our record, we can write the only mass-dependent relation given by
\begin{eqnarray}
\sum_{1\leq i<j<k\leq 4} \sqrt{\frac{{\cal A}_{i}}{4\pi}} \sqrt{\frac{{\cal A}_{j}}{4\pi}} 
\sqrt{\frac{{\cal A}_{k}}{4\pi}} &=& \frac{9M\omega}{\Lambda^2} ~.\label{eq11}
\end{eqnarray}
The entropy \footnote{The entropy in HL gravity is not exactly equal to the area divided by four in addition to that some 
logarithmic term appears because the entropy computed here by using the formula $dM=T dS$ and by assuming the first law of 
thermodynamics is always satisfied \cite{cai9}. In fact, due to some unusual properties of HL gravity the $P-V$ criticality 
also breaks down \cite{mo} in the extended phase space.} of the BH is given by
\begin{eqnarray}
{\cal S}_{i}  &=& \frac{{\cal A}_{i}}{4} ~.\label{eq5}
\end{eqnarray}
and the BH temperature is
$$
T_{i} = \frac{{\cal X}'(r_{i})}{4\pi}
$$
where
$$
{\cal X}'(r_{i})= 2\omega\left(1-\frac{2\Lambda}{3\omega}\right)r_{i}
$$
$$
-2\omega^2\left(1-\frac{4\Lambda}{3\omega}\right)
\frac{r_{i}^3}{\sqrt{1+2\left(1-\frac{2\Lambda}{3\omega} \right)\omega r_{i}^2+
\left(1-\frac{2\Lambda}{3\omega} \right)^2\omega^2 r_{i}^4}}
$$
\begin{eqnarray}
-\frac{\left[4\Lambda^2r_{i}^4+18\left(1-\frac{2\Lambda}{3\omega} \right)\omega r_{i}^2+9\right]}
{18r_{i}\sqrt{1+2\left(1-\frac{2\Lambda}{3\omega} \right)\omega r_{i}^2+
\left(1-\frac{2\Lambda}{3\omega} \right)^2\omega^2 r_{i}^4}}
~. \label{eq7}
\end{eqnarray} 

We find that the CCI for Ho\v{r}ava Lifshitz BH in AdS Space should read off
\begin{eqnarray}
M  &\geq& \frac{\left[18\omega\left(1-\frac{2\Lambda}{3\omega} \right)\left(\frac{{\cal A}_{i}}{4\pi}\right)
+4\Lambda^2\left(\frac{{\cal A}_{i}}{4\pi}\right)^2+9\right]}
{36\omega\sqrt{\frac{{\cal A}_{i}}{4\pi}}} ~.\label{hl3}
\end{eqnarray}
when the inequality becomes equality it is the relation of mass in terms of area. When $\Lambda=0$, we find the CCI
for KS BH in HL gravity
\begin{eqnarray}
M  &\geq & \frac{1}{2} \sqrt{\frac{{\cal A}_{\pm}}{4\pi}}+\frac{1}{4\omega}\sqrt{\frac{4\pi}{{\cal A}_{\pm}}} ~.\label{hl4}
\end{eqnarray}
where ${\cal A}_{\pm}$ is the area of event horizon and Cauchy horizon in KS BH \cite{plb}. 

Now let us calculate the specific heat to determine the local thermodynamic stability of the BH. The specific heat is 
defined by 
\begin{eqnarray}
C_{i} &=& \frac{\partial{\cal M}}{\partial T_{i}}=
\frac{\frac{\partial{\cal M}}{\partial r_{i}}}{\frac{\partial T_{i}}{\partial r_{i}}} .~\label{c1}
\end{eqnarray}
where, 
\begin{eqnarray}
\frac{\partial{\cal M}}{\partial r_{i}} &=& \frac{12\Lambda^2r_{i}^4+18\omega\left(1-\frac{2\Lambda}{3\omega} \right)r_{i}^2-9}
{36\omega r_{i}^4} .~\label{c2}
\end{eqnarray}
and 
\begin{eqnarray}
\frac{\partial T_{i}}{\partial r_{i}} &=& \frac{{\cal X}''(r)}{4\pi}  .~\label{c3}
\end{eqnarray}
where
$$
{\cal X}''(r)= 2\omega\left(1-\frac{2\Lambda}{3\omega}\right)
$$
$$
-2\omega^2\left(1-\frac{4\Lambda}{3\omega}\right)r_{i}^2\frac{\left[3+4\omega\left(1-\frac{2\Lambda}{3\omega} \right)r_{i}^2+
\omega^2 \left(1-\frac{2\Lambda}{3\omega} \right)^2 r_{i}^4\right]}
{\left[1+2\omega\left(1-\frac{2\Lambda}{3\omega} \right) r_{i}^2+
\omega^2 \left(1-\frac{2\Lambda}{3\omega} \right)^2r_{i}^4\right]^{\frac{3}{2}}}
$$
$$
-\frac{\Upsilon (r)}
{18r_{i}^2\left[1+2\omega\left(1-\frac{2\Lambda}{3\omega} \right) r_{i}^2+
\omega^2 \left(1-\frac{2\Lambda}{3\omega} \right)^2r_{i}^4\right]^{\frac{3}{2}}}
$$
where 
$$
\Upsilon (r) =4\omega^2 \Lambda^2 \left(1-\frac{2\Lambda}{3\omega} \right)^2r_{i}^8-
18\omega^3 \left(1-\frac{2\Lambda}{3\omega} \right)^3r_{i}^6+
16\omega \Lambda^2 \left(1-\frac{2\Lambda}{3\omega} \right)r_{i}^6
$$
\begin{eqnarray}
-27\omega^2 \left(1-\frac{2\Lambda}{3\omega} \right)^2r_{i}^4+12\Lambda^2r_{i}^4-
18\omega \left(1-\frac{2\Lambda}{3\omega} \right)r_{i}^2-9   .~\label{c4}
\end{eqnarray}
The local thermodynamic stability requires that $C_{i}>0$ and the second order phase transition occurs at 
${\cal X}''(r)=0$ that means the specific heat diverges at that point.

Now if we want to study the thermodynamic behaviour of the  HL gravity with ADS space  in the extended 
phase space by considering the cosmological constant as a thermodynamic pressure i.e. $\Lambda=-\frac{3}{\ell^2}=-8\pi P$
\footnote{From Eq. (\ref{hlc2}), it follows that the radius of curvature of the asymtotically AdS region is given in terms of 
the following effective cosmological constant i.e. $\Lambda_{eff}=\omega-\frac{2}{3} \Lambda-\sqrt{\omega 
\left(\omega-\frac{4}{3} \Lambda \right)}$. It is convenient to define $\Lambda=-8\pi P$ rather $\Lambda_{eff}=-8\pi P$.} and 
its conjugate quantity as a thermodynamic volume i.e. $V_{i}=\frac{4}{3} \pi r_{i}^3$ then the ADM mass of an AdS BH may be 
treated as the enthalpy of ADS space-time i.e. $M=H=U+PV$. Where $U$ is the thermal energy of the system \cite{kastor09}. 
Therefore one should write the BH equation of state in the extended phase space as 
$$
\left[36 \omega^2 r_{i}^8 \left(1+\frac{16\pi P}{3\omega} \right)^4- 65536 \pi^2 r_{i}^8 P^4 \right] +
\left( 2592 \omega^3 r_{i}^6- 5184 \pi T_{i}\omega^3 r_{i}^7\right) \left(1+\frac{16\pi P}{3\omega}\right)^3 
$$
$$
-\left[ 9216 \pi^2 \omega^2 r_{i}^6 \left(1+\frac{16\pi P}{3\omega}\right) +18432 \pi^2 \omega^2 r_{i}^8 
\left(1+\frac{32 \pi P}{3\omega}\right)\right] P^2
$$
$$
+\left[1296 \omega^2 r_{i}^4- 10368  \pi T_{i}\omega^2 r_{i}^5 +5184 \pi^2 T_{i}^2 \omega^2 r_{i}^6 \right]  
\left(1+\frac{16\pi P}{3\omega}\right)^2
$$
$$
-\left[ 1296 \omega^4 r_{i}^8 \left(1+\frac{32 \pi P}{3\omega}\right)^2
+ 324 \omega^2 r_{i}^4 \left(1+\frac{16\pi P}{3\omega}\right)^2+4608 \pi^2 r_{i}^4 P^2\right]
$$
$$
-1296 \omega^3 r_{i}^6 \left(1+\frac{16\pi P}{3\omega}\right) \left(1+\frac{32\pi P}{3\omega}\right) +
\left(10368 \pi^2 T_{i}^2 \omega r_{i}^4 - 5184 \pi T_{i}\omega r_{i}^3 \right) \left(1+\frac{16\pi P}{3\omega}\right)
$$
\begin{eqnarray}
-\left[18 \omega r_{i}^2 \left(1+\frac{16\pi P}{3\omega}\right) + 648 \omega^2 r_{i}^4\left(1+\frac{32\pi P}{3\omega}\right)\right]
+5184 \pi^2 T_{i}^2  r_{i}^2-36 &=& 0   .~\label{c5}
\end{eqnarray}
From this BH equation of state it is not so trivial task to find the critical constants by imposing the condition at the 
point of inflection point due to to the quartic nature of $P$.

In the extended phase space the mass parameter becomes 
\begin{eqnarray}
M  &=& \frac{256 \pi^2 P^2r_{i}^4+18\omega\left(1+\frac{16\pi P}{3\omega}\right)r_{i}^2+9}{36\omega r_{i}} ~.\label{eqm}
\end{eqnarray}
The thermodynamic volume is defined in the extended phase space as
\begin{eqnarray}
V &=& \left(\frac{\partial M}{\partial P}\right)_{S}  ~.\label{eqv}
\end{eqnarray}
and it is found to be for HL BH in ADS space:
\begin{eqnarray}
V &=& \frac{4}{3}\pi r_{i}^3 \left(\frac{32\pi P}{3\omega}+\frac{2}{\omega r_{i}^2}\right) 
~.\label{eqv1}
\end{eqnarray}
It is quite interesting that the thermodynamic volume for HL BH in ADS space does not satisfied the formula   
\begin{eqnarray}
V_{i} & = & \frac{4}{3}\pi r_{i}^3  ~.\label{eqv2}
\end{eqnarray}
that means HL gravity in ADS space violates the naive geometric volume formula. This is the counter example 
\cite{brena,mo,pparxiv}
of any spherically symmetric BH that the thermodynamic volume is 
\begin{eqnarray}
V_{i} & \neq & \frac{4}{3}\pi r_{i}^3  ~.\label{eqv3}
\end{eqnarray}
But it has been shown that an another interesting feature of the thermodynamic volume is called 
\emph{Reverse Isoperimetric Inequality} \cite{cvetic11} which is  defined as 
\begin{eqnarray}
 {\cal R} &=& \left(\frac{3V}{4\pi} \right)^{\frac{1}{3}} 
 \left(\frac{4\pi}{{\cal A}} \right)^{\frac{1}{2}}  ~.\label{rr}
\end{eqnarray}
and it is calculated to be for this BH:
\begin{eqnarray}
 {\cal R}_{i} &=&  \left(\frac{32\pi P}{3\omega}+\frac{2}{\omega r_{i}^2}\right)^{\frac{1}{3}} >1 ~.\label{rr1}
\end{eqnarray}
It does satisfied for this BH.   

Finally, we compute the Gibbs free energy for this ADS BH and it is found to be
\begin{eqnarray}
 G &=& H-T_{i}S_{i}
 =\frac{256 \pi^2 P^2r_{i}^4+18\omega\left(1+\frac{16\pi P}{3\omega}\right)r_{i}^2+9}{36\omega r_{i}}-r_{i}^2 
 \frac{{\cal X}'(r_{i})}{4}~. \label{gfe}
\end{eqnarray}
It indicates that $G$ depends upon both the thermodynamic pressure and coupling constant.

\section{Area products for BH solutions in massive gravity:}
In this section we would like to derive the thermodynamic product relations particularly area product relation for 
BH solutions in massive gravity \cite{vegh}. The action for BH solutions in massive gravity in $(n+2)$ dimension 
is given by 
\begin{eqnarray}
{\mathcal {I}} &=& \frac{1}{16\pi} \int {\sqrt {-g}}\, {d}^{n+2}x \left[R+\frac{n(n+1)}{\ell^2}-\frac{1}{4} 
F_{\mu\nu} F^{\mu\nu} + m^2 \sum_{i=1}^{4}c_{i} {\cal U}_{i}(g,f) \right] ~.\label{ac}
\end{eqnarray}
where $m$, $\ell$ and $f$ are  graviton mass,  ADS radius and reference metric respectively. The reference metric is defined 
by $f_{\mu\nu}dx^{\mu}dx^{\nu}=h_{ij}dx^{i}dx^{j}$ and $h_{ij}dx^{i}dx^{j}$ represents line element for $n=2$ dimensional 
Einstein space with constant curvature $n(n-1)k$. Where $k=-1, 0, 1$ are the hyperbolic, planar and spherical topology 
of the BH horizons respectively. The field tensor is given by $F_{\mu\nu}=\partial_{\mu}A_{\nu}-\partial_{\nu}A_{\mu}$. 
$c_{i}$ and ${\cal U}_{i}$ are constants and symmetric polynomials of the eigen values of $(n+2)\times (n+2)$ matrix 
${\cal K}_{\nu}^{\mu}\equiv \sqrt{g^{\mu\beta}f_{\beta \nu}}$, i.e., 
\begin{eqnarray}
{\cal U}_{1} &=& [{\cal K}], \nonumber\\
{\cal U}_{2} &=& [{\cal K}]^2-[{\cal K}^2], \nonumber\\
{\cal U}_{3} &=& [{\cal K}]^3-3[{\cal K}][{\cal K}^2]+2[{\cal K}^3], \nonumber\\
{\cal U}_{4} &=& [{\cal K}]^4-6[{\cal K}]^2[{\cal K}^2]+8[{\cal K}][{\cal K}^3]+3[{\cal K}^2]^2-6[{\cal K}^4]
 ~. \label{mge}
\end{eqnarray}
More details can be found in Ref. \cite{cai15}. So we do not repeat here. Finally, we obtain the static BH solutions in 
massive gravity as found in \cite{cai15} like the form as 
\begin{eqnarray}
ds^2 &=& -{\cal F}(r)dt^{2}+\frac{dr^{2}}{{\cal F}(r)}+r^{2} h_{ij} dx^{i}dx^{j} ~.\label{mge1}
\end{eqnarray}
where the ${\cal F}(r)$ is given by 
$$
{\cal F}(r) = k+\frac{r^2}{\ell^2}-\frac{M}{r^{n-1}}+\frac{Q^2}{2n(n-1)r^{2(n-1)}}+\frac{c_{0}c_{1}m^2}{n}r
$$
\begin{eqnarray}
+c_{0}^2c_{2}m^2+\frac{(n-1)c_{0}^3c_{3}m^2}{r}+\frac{(n-1(n-2)c_{0}^2c_{4}m^2}{r^2}~.\label{mge2}
\end{eqnarray}
with the chemical potential at infinity is given by
\begin{eqnarray}
\mu &=& \frac{Q}{(n-1)r_{+}^{n-1}}~.\label{mge3}
\end{eqnarray}
Since in this work we are restricted in the four dimensional case thus we choose the parameter $c_{3}=c_{4}=0$. $M$, $Q$ and 
$c_{0}$ are BH mass, BH charge and a positive constant respectively. For convenient, we have set $\frac{c_{0}c_{1}m^2}{2}=a$ and 
$c_{0}^2c_{2}m^2=b$ respectively. These are remanent graviton mass dependent parameters.  Also we have set the parameter $k=1$ 
for BH case. Therfore the function ${\cal F}(r)$ becomes \cite{zeng} 
\begin{eqnarray}
{\cal F}(r) &=&  1-\frac{2M}{r}+\frac{r^2}{\ell^2}+\frac{Q^2}{4r^2}+ar+b ~.\label{mge4}
\end{eqnarray}
The BH horizons correspond to ${\cal F}(r)=0$:
\begin{eqnarray}
4r^4+4a\ell^2r^3+4(1+b) \ell^2 r^2-8M\ell^2 r+Q^2\ell^2 &=& 0 ~.\label{mge5}
\end{eqnarray}
To finding the roots we apply the Vieta's theorem:
\begin{eqnarray}
\sum_{i=1}^{4} r_{i} &=& -a\ell^2 ~.\label{meq1}\\
\sum_{1\leq i<j\leq 4} r_{i}r_{j} &=& (1+b) \ell^2 ~.\label{meq2}\\
\sum_{1\leq i<j<k\leq 4} r_{i}r_{j} r_{k} &=& 2M\ell^2 ~.\label{meq3}\\
\sum_{1\leq i<j<k<l\leq 4} r_{i}r_{j} r_{k}r_{l} &=& \frac{Q^2\ell^2}{4} ~.\label{meq4}
\end{eqnarray}

Case I: 

Since we are interested to finding the mass-independent area product formula in terms of two physical horizons 
thus by eliminating mass parameter from the above Eqs. we have found 
\begin{eqnarray}
\frac{\frac{Q^2\ell^2}{4}}{r_{1}r_{2}}-a\ell^2(r_{1}+r_{2})-\left(r_{1}^2+r_{2}^2 +r_{1}r_{2}\right) &=& (1+b) \ell^2
~.\label{meq8}
\end{eqnarray}
If we are working with area of the BH horizons i.e. $A_{i}=4\pi r_{i}^2 (i=1,2)$ then we should find the following 
mass-independent area functional relation in terms of two BH physical horizons:
\begin{eqnarray}
\frac{\pi Q^2 \ell^2}{\sqrt{{\cal A}_{1}{\cal A}_{2}}}-a\ell^2\left[\sqrt{\frac{{\cal A}_{1}}{4\pi}}+\sqrt{\frac{{\cal A}_{2}}{4\pi}}
\right]-\frac{\left({\cal A}_{1}+{\cal A}_{2}+\sqrt{{\cal A}_{1}{\cal A}_{2}}\right)}{4\pi} &=& (1+b) \ell^2 ~.\label{meq9}
\end{eqnarray}
Once again the other mass-independent formula becomes
\begin{eqnarray}
\sqrt{{\cal A}_{1}{\cal A}_{2}{\cal A}_{3}{\cal A}_{4}} &=& 4\pi^2 \ell^2 Q^2 ~.\label{meq10}
\end{eqnarray}

Case II:
For our record, we have only single mass dependent area functional relation is given by 
\begin{eqnarray}
\left[\frac{\pi Q^2 \ell^2}{\sqrt{{\cal A}_{1}{\cal A}_{2}}}-\frac{{\sqrt{{\cal A}_{1}{\cal A}_{2}}}}{4\pi}\right]
\left[\sqrt{\frac{{\cal A}_{1}}{4\pi}}+\sqrt{\frac{{\cal A}_{2}}{4\pi}}\right]-a\ell^2\frac{{\sqrt{{\cal A}_{1}{\cal A}_{2}}}}{4\pi} 
&=& 2M \ell^2 ~.\label{meq11}
\end{eqnarray}

The BH temperature in massive gravity is given by
$$
T_{i} = \frac{{\cal F}'(r_{i})}{4\pi}
$$
where
\begin{eqnarray}
= \frac{12r_{i}^4+8a\ell^2r_{i}^3+4\ell^2(1+b)r_{i}-\ell^2Q^2}{16\pi \ell^2r_{i}^3} ~.\label{teq}
\end{eqnarray}
where $i=1,2$.

The CCI for this BH should read 
\begin{eqnarray}
M  & \geq &  
\sqrt{\frac{{\cal A}_{i}}{16\pi}}+\left(\frac{{\cal A}_{i}}{4\pi\ell^2}\right)\sqrt{\frac{{\cal A}_{i}}{16\pi}}+\frac{a}{2}
+b\sqrt{\frac{{\cal A}_{i}}{16\pi}}  ~.\label{epi}
\end{eqnarray}

\begin{figure}[h]
 \begin{center}
 \subfigure[ ]{
 \includegraphics[width=2.1in,angle=0]{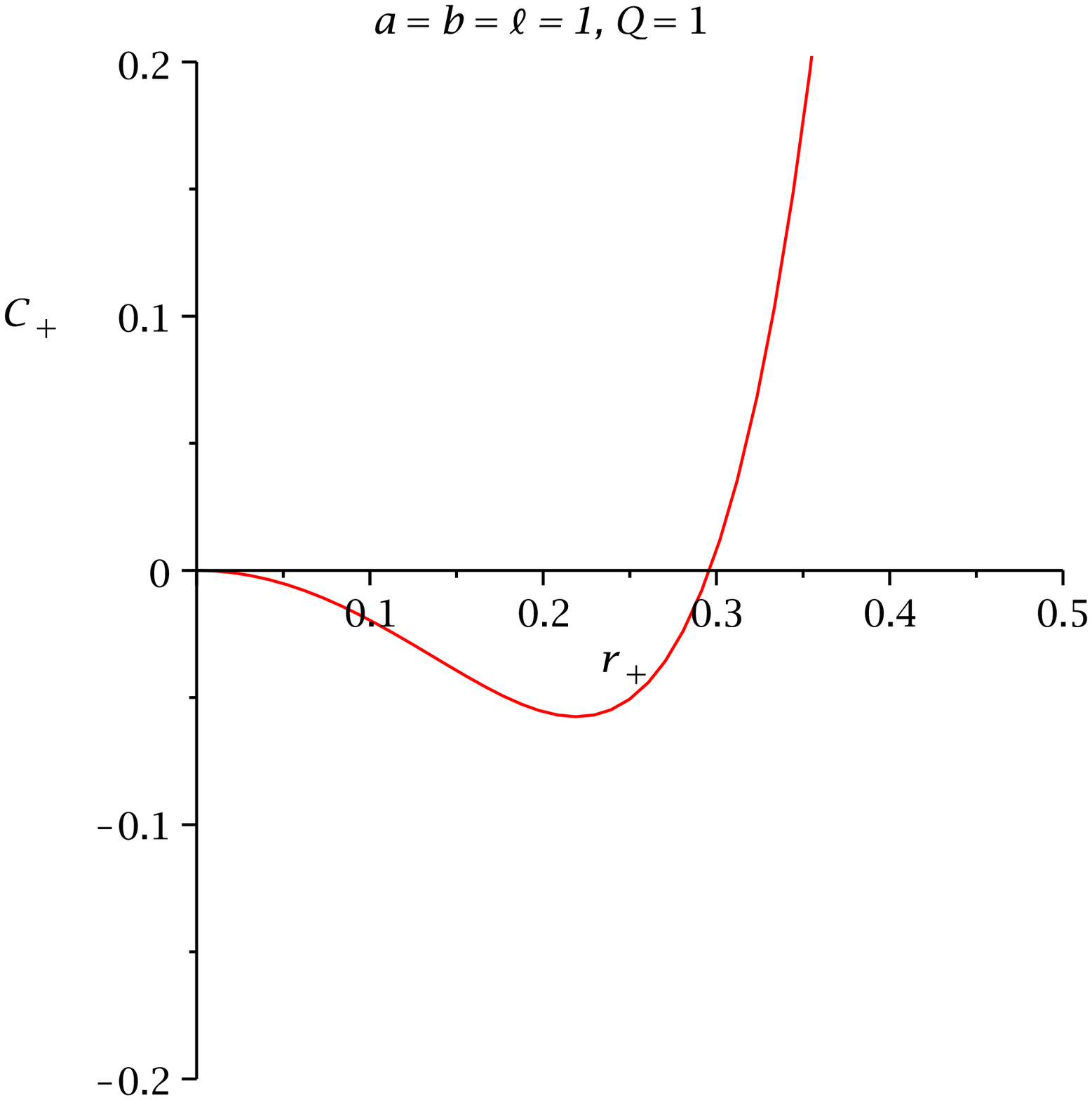}}
 \subfigure[ ]{
 \includegraphics[width=2.1in,angle=0]{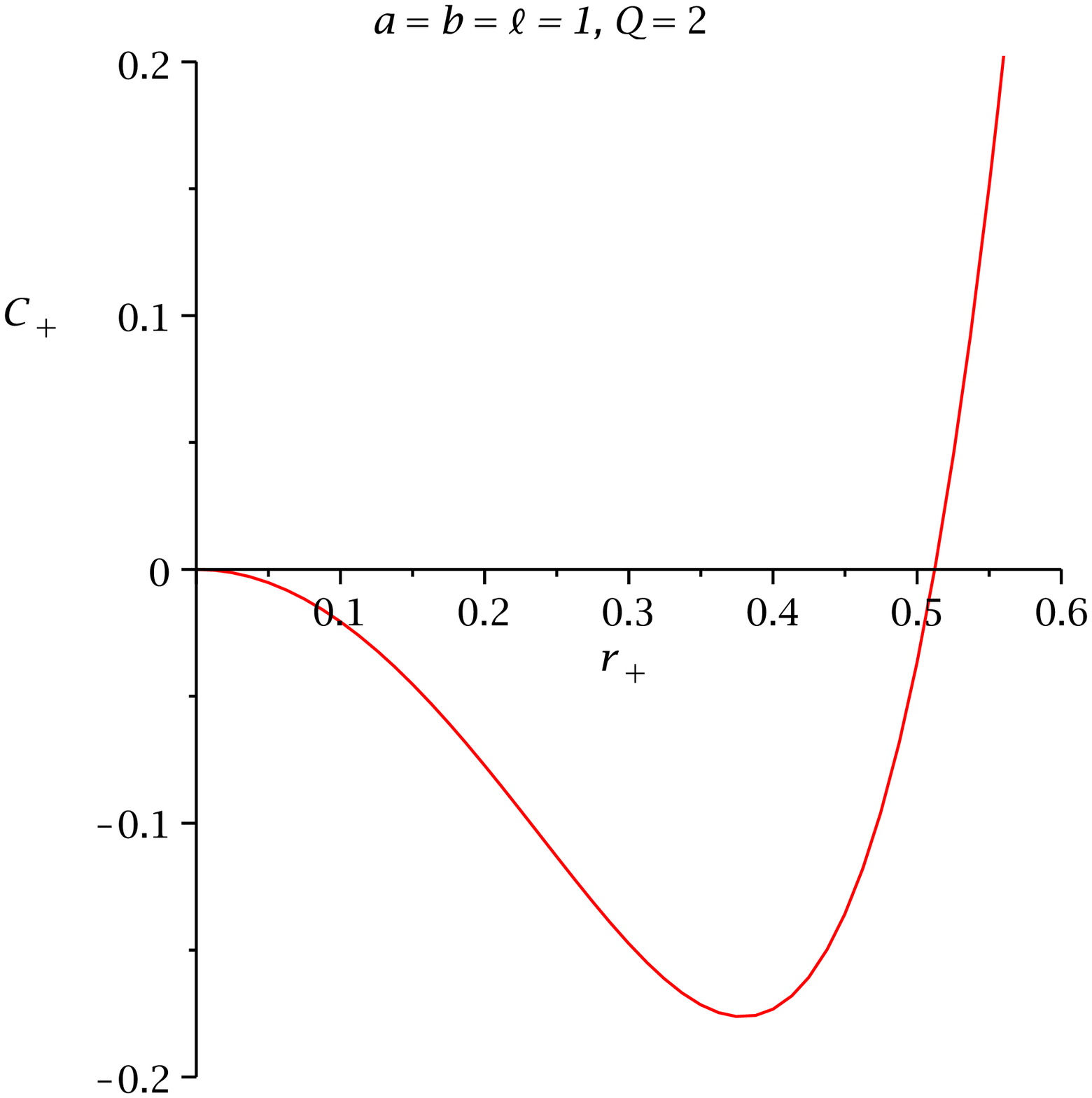}}
 \caption{\label{fgm}\textit{ In this figure, we have plotted the variation of specific heat with horizon radius for massive 
 gravity.  }} 
\end{center}
\end{figure}

Finally the specific heat is given by 
\begin{eqnarray}
C_{i} &=& 2\pi r_{i}^2 \frac{\left[12r_{i}^4+8a\ell^2r_{i}^3+4\ell^2(1+b)r_{i}^2-\ell^2Q^2\right]}
{\left[12r_{i}^4-4\ell^2(1+b)r_{i}^2+3\ell^2Q^2\right]} .~\label{emc}
\end{eqnarray}
The specific heat diverges at 
\begin{eqnarray}
r_{i} &=& \pm \sqrt{\frac{(1+b)\ell^2}{6}\left[1\pm\sqrt{1-\frac{12Q^2}{\ell^2(1+b)^2}} \right]} .~\label{emc1}
\end{eqnarray}
which indicates that the BH in massive gravity showing second order phase transition. It could be seen from the Fig. \ref{fgm}. 

Now if we consider the extended phase space the mass parameter becomes 
\begin{eqnarray}
M &=& \frac{4}{3} \pi P r_{i}^3+\frac{r_{i}}{2}+\frac{Q^2}{8r_{i}}+\frac{a}{2}+\frac{br_{i}}{2} .~\label{emc2}
\end{eqnarray}
The thermodynamic volume is found to be 
\begin{eqnarray}
V_{i} &=& \left(\frac{\partial M}{\partial P}\right)_{S} =\frac{4}{3}\pi r_{i}^3   ~.\label{eqvv}
\end{eqnarray}
Unlike HL gravity with AdS space it does satisfied that the thermodynamic volume is equal to the naive geometric 
volume.

The Gibbs free energy for massive gravity is calculated to be 
\begin{eqnarray}
G_{i} &=& M-T_{i}S_{i}
=\frac{4}{3} \pi P r_{i}^3+\frac{r_{i}}{2}+\frac{Q^2}{8r_{i}}+\frac{a}{2}+\frac{br_{i}}{2}-r_{i}^2 
\frac{{\cal F}'(r_{i})}{4}~. \label{mgg}
\end{eqnarray}

\section{Area products for phantom AdS BHs:}
In this section we would like to compute the thermodynamic product relations of charged phantom spherically symmetric AdS 
BHs \cite{hq}.
The main motivation behind this investigation is come from the fact that phantom fields are exotic fields in theoretical physics.
This phantom field is due to negative energy density. Current observational data \cite{jd} tells us that phantom field can explain 
the acceleration of our universe. Thus it is important to study the thermodynamic properties of phantom fields by calculating 
thermodynamic products particularly area (or entropy ) products.

The action for an AdS BH with phantom charge is given by 
\begin{eqnarray}
{\mathcal {I}} &=& \frac{1}{16\pi} \int {\sqrt {-g}}\, {d}^{4}x \left[R-2\Lambda+2\eta F_{\mu\nu} F^{\mu\nu} \right] 
~.\label{acp}
\end{eqnarray}
where the constant $\eta$ determines the nature of the electro-magnetic (EM) field. For $\eta=1$, we obtain the classical 
EM theory but when $\eta=-1$, one gets the Maxwell field which is \emph{phantom}. The spherically symmetric solution of the above 
action is given by 
\begin{eqnarray}
ds^2=-{\cal V}(r)dt^{2}+\frac{dr^{2}}{{\cal V}(r)}+r^{2}(d\theta^{2}+\sin^{2}\theta d\phi^{2}) 
~.\label{ph}
\end{eqnarray}
where the function ${\cal V}(r)$ is 
\begin{eqnarray}
{\cal V}(r) &=& 1-\frac{2M}{r}-\frac{\Lambda}{3}r^2+\eta\frac{Q^2}{r^2} ~.\label{ph1}
\end{eqnarray}
and,  
\begin{eqnarray}
F &=& \frac{1}{2}F_{\mu\nu} dx^{\mu}\wedge dx^{\nu}= \frac{Q^2}{r^2} dt \wedge dr ~.\label{ph2}
\end{eqnarray}
and also $M$ and $Q$ are represents the mass of the BH and charge of the EM source. When $\eta=-1$, the energy of the EM 
field to the action becomes negative so it could be interpreted as \emph{exotic} matter.

Now setting $-\frac{\Lambda}{3}=\frac{1}{\ell^2}$. Therefore the BH horizons correspond to ${\cal V}(r)=0$:
\begin{eqnarray}
r^4+\ell^2r^2-2M\ell^2 r+\eta \ell^2Q^2 &=& 0 ~.\label{pge}
\end{eqnarray}
Proceeding analogously, applying Vieta's theorem we find
\begin{eqnarray}
\sum_{i=1}^{4} r_{i} &=& 0 ~.\label{peq1}\\
\sum_{1\leq i<j\leq 4} r_{i}r_{j} &=&  \ell^2 ~.\label{peq2}\\
\sum_{1\leq i<j<k\leq 4} r_{i}r_{j} r_{k} &=& 2M\ell^2 ~.\label{peq3}\\
\sum_{1\leq i<j<k<l\leq 4} r_{i}r_{j} r_{k}r_{l} &=& \eta Q^2\ell^2 ~.\label{peq4}
\end{eqnarray}

Case I: 

Again eliminating mass parameter, one finds only single mass-independent relation in terms of 
two physical horizons:
\begin{eqnarray}
\eta \frac{\ell^2 Q^2}{r_{1}r_{2}}-\left(r_{1}^2+r_{2}^2 +r_{1}r_{2}\right) &=& \ell^2 ~.\label{peq8}
\end{eqnarray}

Similarly one can rewrite this formula in terms of area ${\cal A}_{i} = 4\pi r_{i}^2, \, (i=1,2)$:
\begin{eqnarray}
\eta \frac{\left(4\pi \ell Q\right)^2}{\sqrt{{\cal A}_{1}{\cal A}_{2}}}-
\left({\cal A}_{1}+{\cal A}_{2}+\sqrt{{\cal A}_{1}{\cal A}_{2}}\right) &=& 4\pi\ell^2 ~.\label{peq9}
\end{eqnarray}
The other mass-independent formula becomes
\begin{eqnarray}
\sqrt{{\cal A}_{1}{\cal A}_{2}{\cal A}_{3}{\cal A}_{4}} &=& 16 \pi^2 \eta \ell^2 Q^2  ~.\label{peq10}
\end{eqnarray}

Case II:

It may be noted that the only single mass dependent area product relation is
\begin{eqnarray}
\left[\eta \frac{(4\pi \ell Q)^2}{\sqrt{{\cal A}_{1}{\cal A}_{2}}}-{\sqrt{{\cal A}_{1}{\cal A}_{2}}}\right]
\left[\sqrt{\frac{{\cal A}_{1}}{4\pi}}+\sqrt{\frac{{\cal A}_{2}}{4\pi}}\right]-a\ell^2\frac{{\sqrt{{\cal A}_{1}{\cal A}_{2}}}}{4\pi} 
&=& 8\pi M \ell^2 ~.\label{peq11}
\end{eqnarray}

The BH temperature should read 
\begin{eqnarray}
T_{i} &=& \frac{{\cal V}'(r)}{4\pi}= 
\frac{1}{4\pi r_{i}} \left(1+3 \frac{r_{i}^2}{\ell^2}-\eta \frac{Q^2}{r_{i}^2} \right)~. \label{peq7}
\end{eqnarray}
where $i=1,2$.

The CCI for phantom BHs should be
\begin{eqnarray}
M  & \geq &  
\sqrt{\frac{{\cal A}_{i}}{16\pi}}\left(1+\frac{{\cal A}_{i}}{4\pi\ell^2}+\eta \frac{4\pi Q^2}{{\cal A}_{i}}\right) ~.\label{pepi}
\end{eqnarray}

\begin{figure}[h]
 \begin{center}
 \subfigure[ ]{
 \includegraphics[width=2.1in,angle=0]{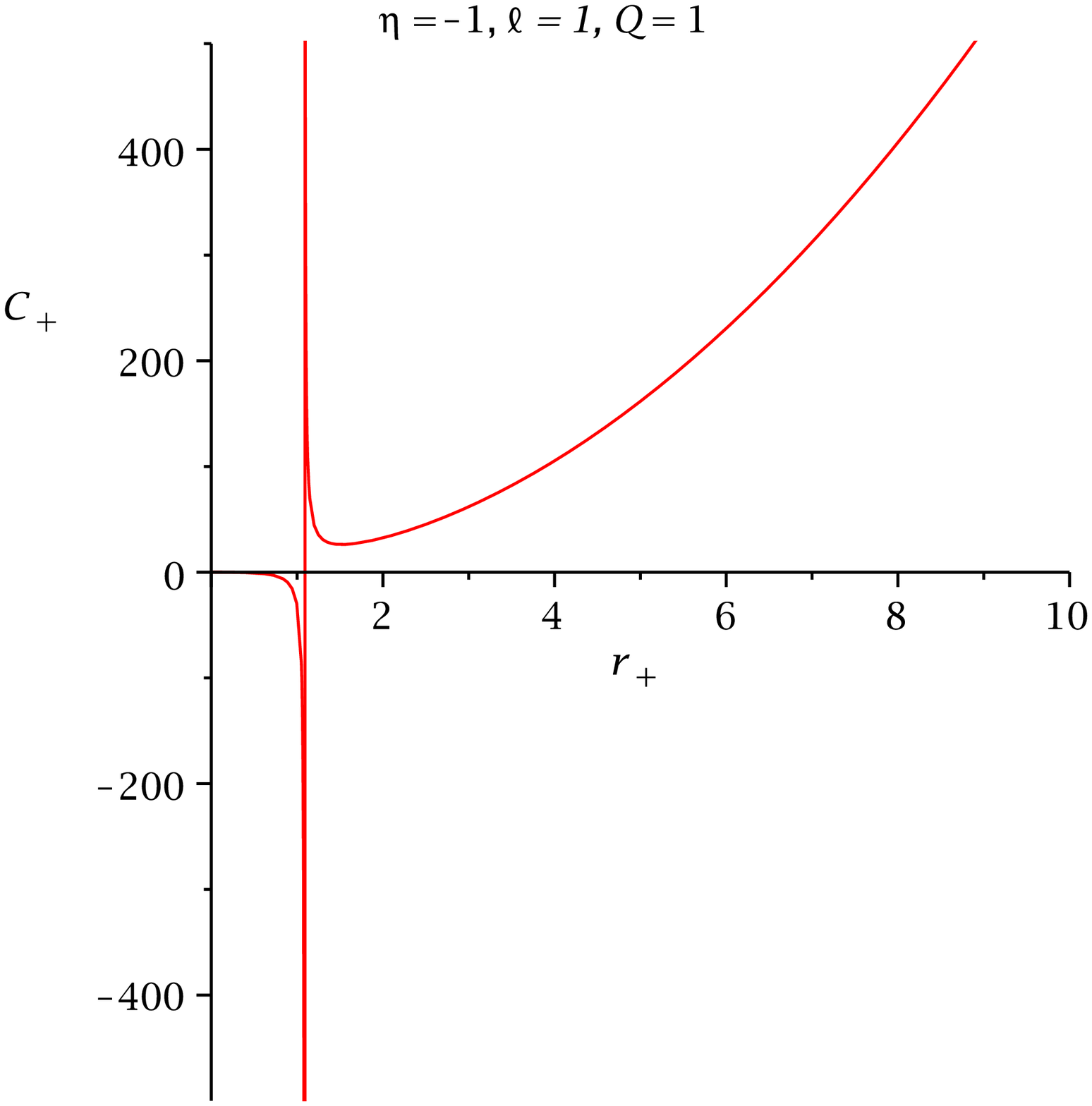}}
 \subfigure[ ]{
 \includegraphics[width=2.1in,angle=0]{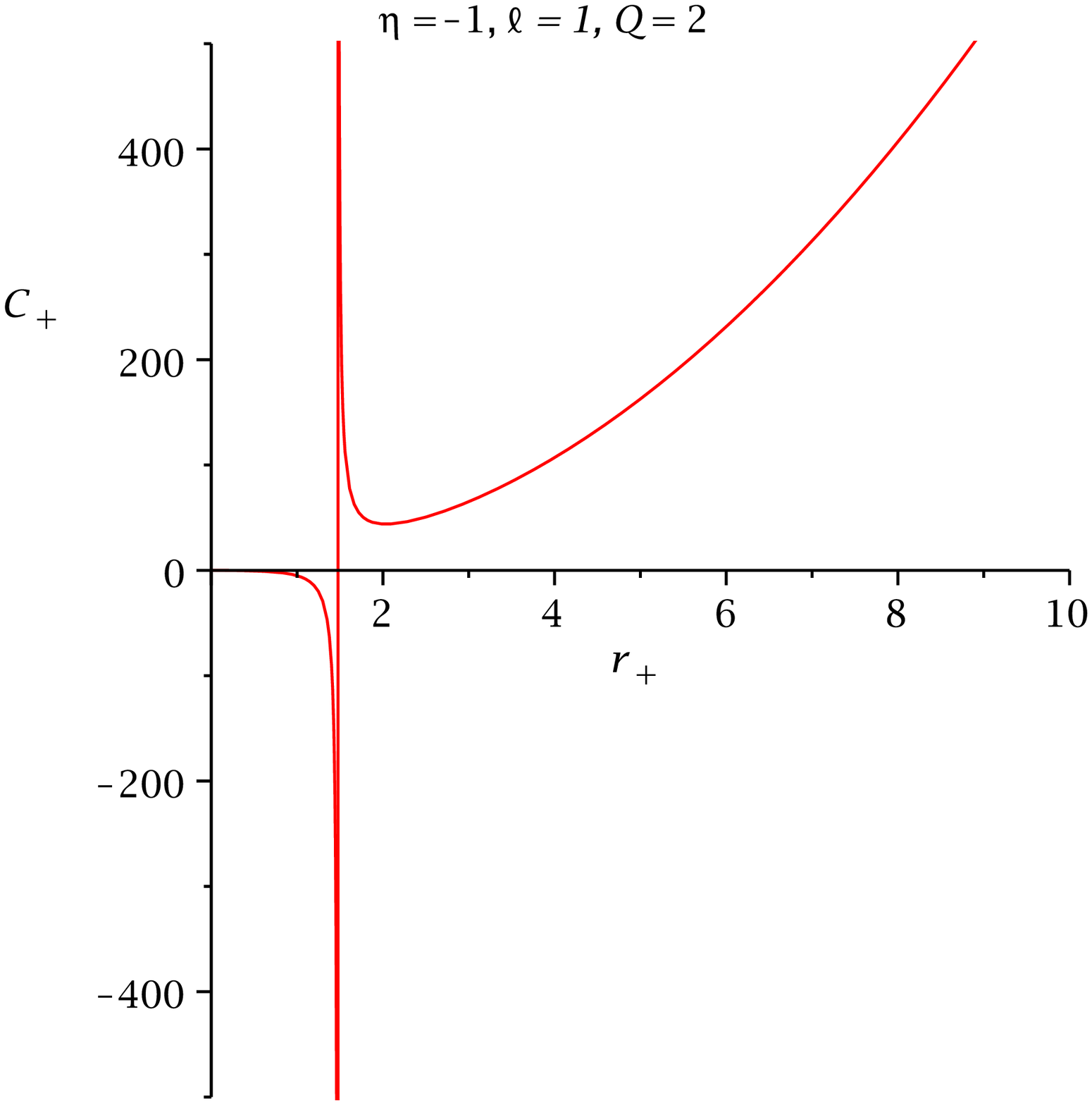}}
 \caption{\label{fgp}\textit{ In this figure, we have plotted the variation of specific heat with horizon radius for 
 phantom AdS BH. }.} 
\end{center}
\end{figure}

The specific heat is calculated to be 
\begin{eqnarray}
C_{i} &=& 2 \pi r_{i}^2 \left[\frac{3\frac{r_{i}^2}{\ell^2}-\eta \frac{Q^2}{r_{i}^2}+1}
{3\frac{r_{i}^2}{\ell^2}+3\eta \frac{Q^2}{r_{i}^2}-1} \right].~\label{prn3}
\end{eqnarray}
The specific heat diverges at 
\begin{eqnarray}
r_{i} &=& \pm \sqrt{\frac{\ell^2}{6}\left[1\pm\sqrt{1-36\eta\frac{Q^2}{\ell^2}} \right]} .~\label{pemc1}
\end{eqnarray}
that indicates the phantom BHs possesses second order phase transition. It could be found from the Fig. \ref{fgp}.

In the extended phase space the mass parameter becomes 
\begin{eqnarray}
M &=& \frac{4}{3} \pi P r_{i}^3+\frac{r_{i}}{2}+\eta \frac{Q^2}{2r_{i}} .~\label{pemc2}
\end{eqnarray}
The thermodynamic volume in this case is found to be 
\begin{eqnarray}
V_{i} &=& \left(\frac{\partial M}{\partial P}\right)_{S} =\frac{4}{3}\pi r_{i}^3   ~.\label{peqvv}
\end{eqnarray}
Unlike HL gravity in AdS space, it indeed do satisfied that the thermodynamic volume is equal to the 
naive geometric volume.

The Gibbs free energy for phantom BH is given by 
\begin{eqnarray}
G_{i} &=& M-T_{i}S_{i}
=\frac{4}{3} \pi P r_{i}^3+\frac{r_{i}}{2}+\eta \frac{Q^2}{2r_{i}}-r_{i}^2 
\frac{{\cal V}'(r_{i})}{4}~. \label{pmgg}
\end{eqnarray}
It may be noted that the product of Gibbs free energy depends upon the mass parameter. 
In the limit $\eta=1$, all the results reduce to for RN-AdS BH and for $\eta=-1$, one obtain the 
results for \emph{phantom} BHs.  

\section{Area products for Super-Entropic BHs:}
This section deals with a new class of spinning AdS BHs having non-compact event horizons of finite area in four 
dimensions and they are the solutions of gauged supergravity \cite{ag}. Additionaly, their entropy exceeds the maximal bound 
followed by the conjecture \emph {Reverse Isoperimetric Inequality}. This type of BHs are called super-entropic BH 
\cite{hmk,dk,hmk1}.

The metric for super-entropic BH \cite{hmk} (in units where $G=c=1$) is given by 
\begin{eqnarray}
ds^2 &=& -\frac{\Delta}{\rho^2} \, \left[dt-\ell sin^2\theta d\psi \right]^2+\frac{\sin^4\theta}{\rho^2} \,
\left[\ell dt-(r^2+\ell^2) \,d\psi\right]^2
+\rho^2 \, \left[\frac{dr^2}{\Delta}+\frac{d\theta^2}{sin^2\theta}\right] ~.\label{se}
\end{eqnarray}
where
\begin{eqnarray}
\rho^2 &\equiv& r^2+\ell^2 cos^2\theta \\
\Delta &\equiv& \left(\ell+\frac{r^2}{\ell}\right)^2-2mr+q^2   ~.\label{se1}
\end{eqnarray}

The horizon location can be determined by the condition $\Delta (r)=0$
\begin{eqnarray}
r^4+2\ell^2r^2-2m\ell^2 r+ \ell^2 (\ell^2+q^2) &=& 0 ~.\label{se3}
\end{eqnarray}
Applying Vieta's rule we get
\begin{eqnarray}
\sum_{i=1}^{4} r_{i} &=& 0 ~.\label{seq1}\\
\sum_{1\leq i<j\leq 4} r_{i}r_{j} &=& 2 \ell^2 ~.\label{seq2}\\
\sum_{1\leq i<j<k\leq 4} r_{i}r_{j} r_{k} &=& 2m\ell^2 ~.\label{seq3}\\
\sum_{1\leq i<j<k<l\leq 4} r_{i}r_{j} r_{k}r_{l} &=& \ell^2 (\ell^2+q^2) ~.\label{seq4}
\end{eqnarray}
The relevent thermodynamic quantities of the super-entropic BH are 
\begin{eqnarray}
M &=& \frac{m\mu}{2\pi}, \,\, J = M \ell,\,\, \Omega_{i}=\frac{\ell}{r_{i}^2+\ell^2} ~.\label{seq5}\\
{\cal S}_{i} &=& \frac{{\cal A}_{i}}{4}=\frac{\mu}{2} \left(r_{i}^2+\ell^2 \right), \,\, \Phi_{i}=\frac{qr_{i}}{r_{i}^2+\ell^2}, 
\,\, Q=\frac{\mu q}{2\pi} ~.\label{seq6}
\end{eqnarray}
where $\mu$ is dimensionless parameter and $i=1,2$.

Case I: 

Now eliminating the mass parameter, one obtains the mass-independent relation in terms of 
two physical horizons:
\begin{eqnarray}
\frac{\ell^2 (q^2+\ell^2)}{r_{1}r_{2}}-\left(r_{1}^2+r_{2}^2 +r_{1}r_{2}\right) &=& 2\ell^2 ~.\label{seq8}
\end{eqnarray}

Similarly, one obtains this formula in terms of area ${\cal A}_{i} = 2\mu (r_{i}^2+\ell^2), \, (i=1,2)$:
\begin{eqnarray}
\frac{\ell^2 (q^2+\ell^2)}{\sqrt{\frac{{\cal A}_{1}}{2\mu}-\ell^2}\sqrt{\frac{{\cal A}_{2}}{2\mu}-\ell^2}}-
\left[\left(\frac{{\cal A}_{1}}{2\mu}+\frac{{\cal A}_{2}}{2\mu}\right)-2\ell^2+
\sqrt{\frac{{\cal A}_{1}}{2\mu}-\ell^2}\sqrt{\frac{{\cal A}_{2}}{2\mu}-\ell^2} \right] &=& 2\ell^2 ~.\label{seq9}
\end{eqnarray}
The other mass-independent formula for super-entropic BHs is 
\begin{eqnarray}
\sqrt{\frac{{\cal A}_{1}}{2\mu}-\ell^2}\sqrt{\frac{{\cal A}_{2}}{2\mu}-\ell^2}
\sqrt{\frac{{\cal A}_{3}}{2\mu}-\ell^2}\sqrt{\frac{{\cal A}_{4}}{2\mu}-\ell^2} &=&  \ell^2 (q^2+\ell^2) ~.\label{seq10}
\end{eqnarray}

Case II:

The mass dependent area product formula for super-entropic BH should read
\begin{eqnarray}
\left[\frac{\ell^2 (q^2+\ell^2)}{\sqrt{\frac{{\cal A}_{1}}{2\mu}-\ell^2}\sqrt{\frac{{\cal A}_{2}}{2\mu}-\ell^2}}-
\sqrt{\frac{{\cal A}_{1}}{2\mu}-\ell^2}\sqrt{\frac{{\cal A}_{2}}{2\mu}-\ell^2}\right]
\left[\sqrt{\frac{{\cal A}_{1}}{2\mu}-\ell^2}+\sqrt{\frac{{\cal A}_{2}}{2\mu}-\ell^2} \right]
&=& 2m\ell^2 ~.\label{seq11}
\end{eqnarray}

The BH temperature  should read 
\begin{eqnarray}
T_{i} &=&  \frac{1}{4\pi r_{i}} \left(3 \frac{r_{i}^2}{\ell^2}-1- \frac{q^2}{r_{i}^2+\ell^2} \right)~. \label{seq7}
\end{eqnarray}
where $i=1,2$.

The CCI for this BH is calculated to be
\begin{eqnarray}
m  & \geq & \frac{\left[\ell^2 q^2+\left( \sqrt{\frac{{\cal A}_{i}}{2\mu}-\ell^2} +\ell^2 \right)^2 \right]}
{2\ell^2 \sqrt{\frac{{\cal A}_{i}}{2\mu}-\ell^2}}  ~.\label{spi}
\end{eqnarray}

\begin{figure}[h]
 \begin{center}
 \subfigure[ ]{
 \includegraphics[width=2.1in,angle=0]{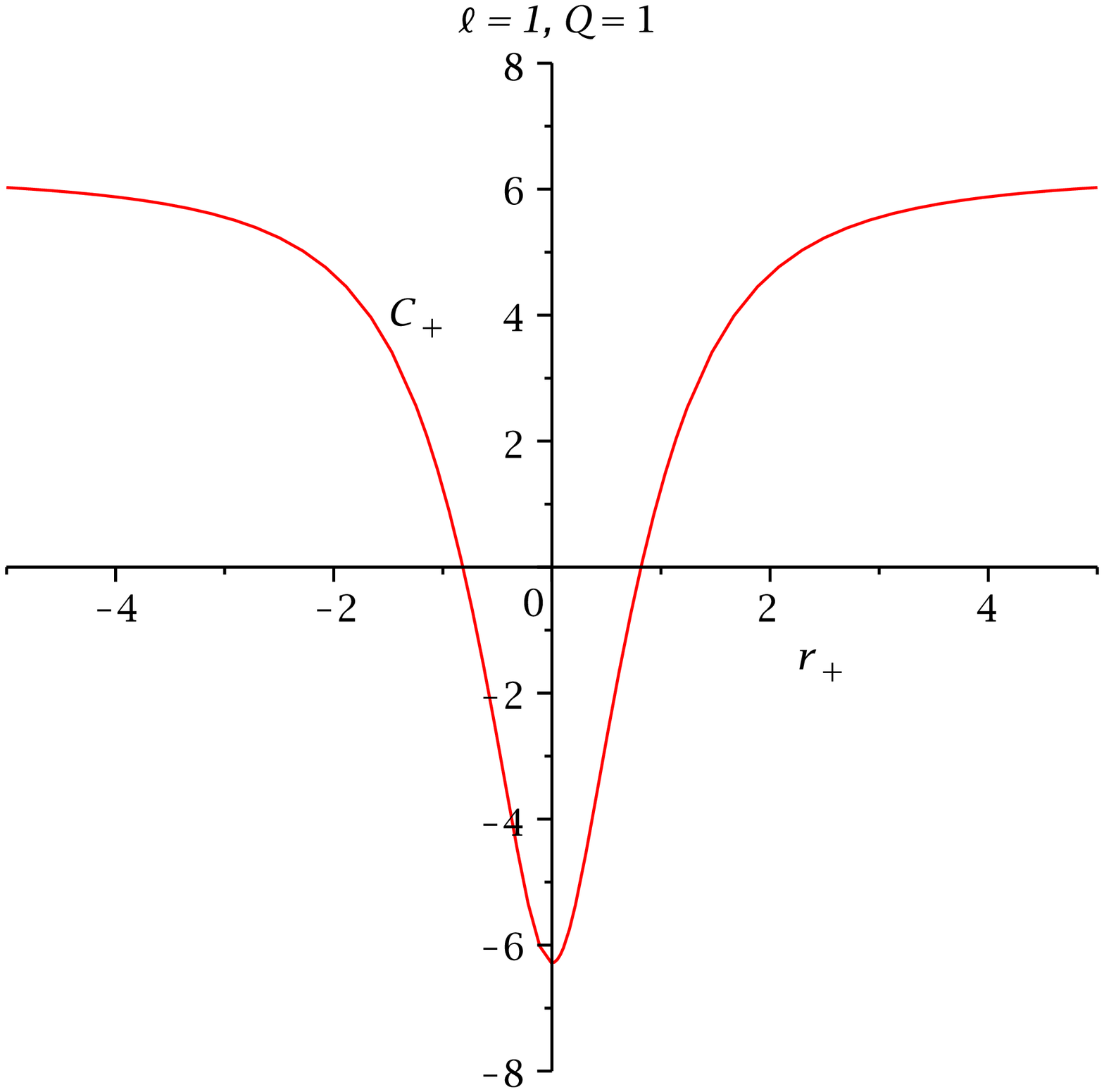}}
 \subfigure[ ]{
 \includegraphics[width=2.1in,angle=0]{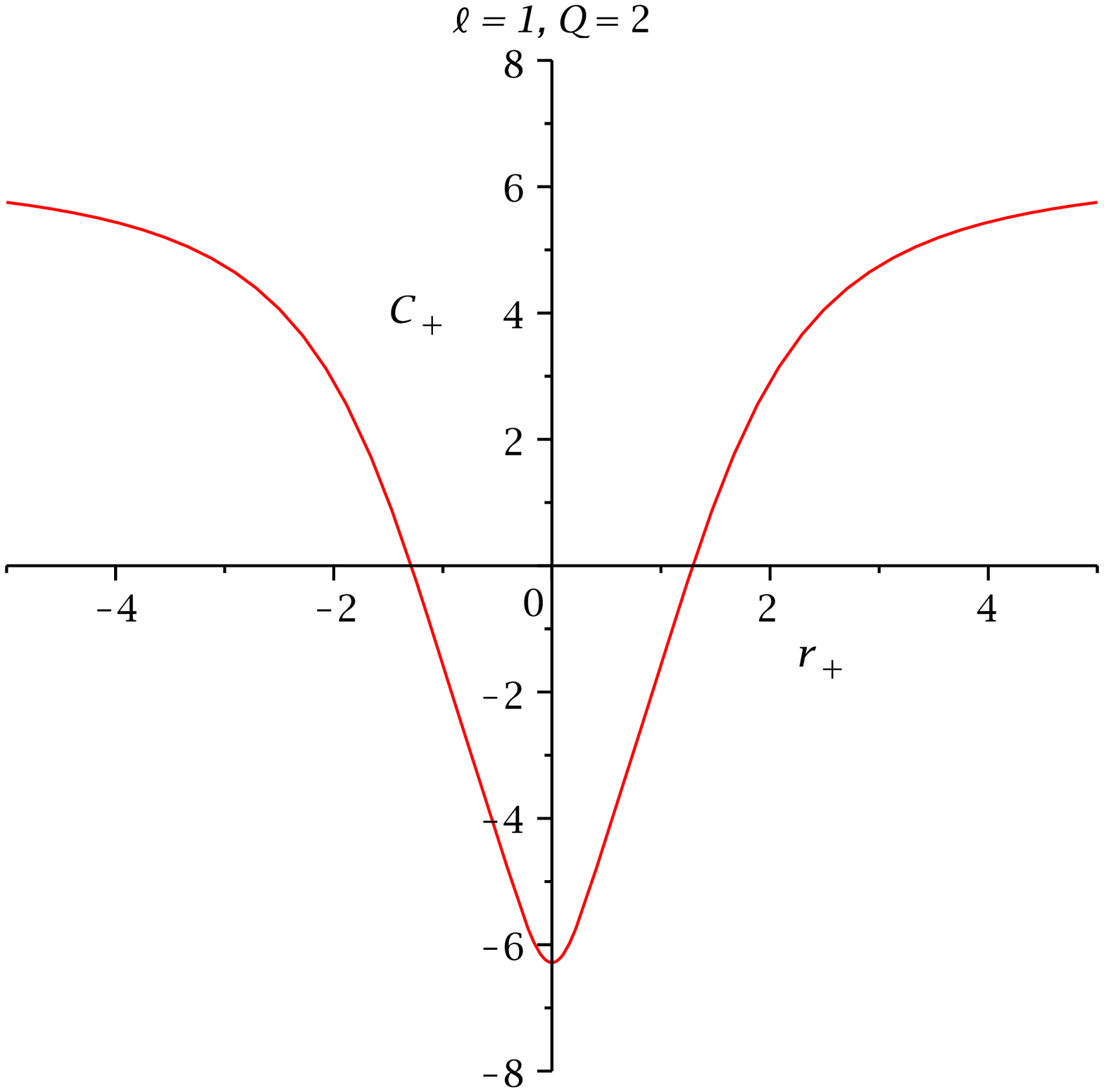}}
 \caption{\label{fgs}\textit{ In this figure, we have plotted the variation of specific heat with horizon radius for 
 super-entropic BH.}} 
\end{center}
\end{figure}

The specific heat is found to be  
\begin{eqnarray}
C_{i} &=& \frac{2 \pi}{\ell^2} \left[ \frac{\left(r_{i}^2+\ell^2 \right) \left(3r_{i}^2-\ell^2\right)-q^2\ell^2}
{1+3 \frac{r_{i}^2}{\ell^2}+q^2 \frac{\left(3r_{i}^2+\ell^2\right)}{\left(r_{i}^2+\ell^2 \right)^2}}\right]
.~\label{srn3}
\end{eqnarray}
The specific heat diverges for super-entropic BH at the point by solving the following equation 
\begin{eqnarray}
3r_{i}^6+7\ell^2r_{i}^4+\ell^2\left(5\ell^2+3q^2\right)r_{i}^2+\ell^4(q^2+\ell^2) &=& 0 .~\label{pss11}
\end{eqnarray}
It seems that this equation has no real solution at all and from the Fig. \ref{fgs}, it is clear that there 
are in fact no second order phase transition occurs for four dimensional super-entropic BHs.

The thermodynamic volume for super-entropic BH is calculated in \cite{hmk} 
\begin{eqnarray}
V_{i} &=& \frac{2}{3}\mu r_{i}\left(r_{i}^2+\ell^2 \right)    ~.\label{seqvv}
\end{eqnarray}
which is reminiscent of the naive geometric volume. It follows from the equation is that it is independent of 
the BH charge. It is also important to note that why this BH is called super-entropic that can calculated in 
\cite{hmk} for ${\cal H}^{+}$:
\begin{eqnarray}
 {\cal R}_{i} &=&  \left(\frac{r_{i}^2}{r_{i}^2+\ell^2}\right)^{\frac{1}{6}} < 1 ~.\label{srr1}
\end{eqnarray}
that means it does not satisfied the condition ${\cal R}_{i}>1$, this is why it is called super-entropic BH. 

If we consider the extended phase space, the mass parameter becomes 
\begin{eqnarray}
m &=& \frac{1}{2r_{i}} \left[q^2+\frac{8\pi P}{3} \left(r_{i}^2+\frac{3}{8\pi P} \right)^2 \right] .~\label{semc2}
\end{eqnarray}

Therefore the Gibbs free energy is computed to be  
\begin{eqnarray}
G_{i} &=& m-T_{i}S_{i} \nonumber\\
&=& \frac{1}{2r_{i}} \left[q^2+\frac{8\pi P}{3} \left(r_{i}^2+\frac{3}{8\pi P} \right)^2 \right]
-\frac{\mu}{8\pi r_{i}} \left(r_{i}^2+\frac{3}{8 \pi P} \right)
\left[8\pi P r_{i}^2-1-\frac{q^2}{\left(r_{i}^2+\frac{3}{8 \pi P} \right)} \right]~. \nonumber \\
\label{smgg}
\end{eqnarray}
It indicates that the product of Gibbs free energy does depend on the mass parameter, so it is not a mass-independent 
quantity.

\section{Discussion:}

We studied the intriguing thermodynamic properties for different class of  asymtotically BH solutions in AdS space in Einstein 
gravity as well as certain modified theories of gravity. We computed their area (or entropy) product relations that are 
mass-independent and which also relates the different BH parameters as well as area of ${\cal H}^{\pm}$. These mass-independent 
thermodynamic  products for multi-horizon BHs have interesing implications in BH thermodynamics. First, these relations signals 
they could turn out to be an universal quantity. Second, they  providig some insight into  the microscopic origin of non-extremal 
BH entropy (both inner and outer) which is an outstanding problems in quantum gravity. We have also studied the local thermodynamic 
stability by computing the specific heat. We determined the condition under which some BHs possesses second order phase transition. 
Furthermore, we have derived the important inequality in general relativity which is so called CCI which relates the mass and 
area of the BH. Finally, we computed the Gibb's free energy for each BHs.


\end{document}